\documentstyle[11pt]{article}
\def\th{\theta} 
\def\pa{\partial}

\def\a{\alpha}
\def\b{\beta}
\def\d{\delta}

\def\l{\lambda} \def\L{\Lambda}
\def\m{\mu}

\def\o{\omega}

\newcommand{\cO}{{\cal O}}
\def\be{\begin{equation}}
\def\ee{\end{equation}}

\def\bea{\begin{eqnarray}}
\def\eea{\end{eqnarray}}

\begin{document}

\begin{flushright} BRX TH-579 \end{flushright}


\begin{center}
{\Large\bf Confinement/Deconfinement Transition of Large $N$\\
Gauge Theories in Perturbation Theory with $N_f$ \\
Fundamentals: $N_f/N$ Finite}

{\large Howard J.\ Schnitzer}\footnote{email:
schnitzr@brandeis.edu\\\hspace*{.2in}Supported in part by the DOE
under
grant DE-FG02-92ER40706}\\

Theoretical  Physics Group\\
Martin Fisher School of Physics, Brandeis University\\
Waltham, MA 02454

\end{center}


\renewcommand{\theequation}{1.\arabic{equation}}
\setcounter{equation}{0}

\begin{abstract}
{\normalsize In hep-th/0402219 we considered large $N$
zero-coupling $d$-dimensional $U(N)$ gauge theories, with $N_f$
matter fields in the fundamental representation on a compact
spatial manifold $S^{d-1} \times$ time, with $N_f/N$ finite.  This
class of theories undergo a 3$^{\rm rd}$ order deconfinement
transition. As a consequence it was proposed that the dual string
theory has a 3$^{\rm rd}$ order phase-transition to a black hole
at high temperature.

In this paper we argue that the same conclusions are valid for such
theories at any finite order in perturbation theory in the 't~Hooft
coupling $\lambda$ and $N_f/N$. It is conjectured that this
continues to hold at strong coupling and finite value of
$N_f/N<1/\lambda$, which suggests that the supergravity
approximation to the dual string theory has a 3$^{\rm rd}$ order
thermal phase transition to a large black hole.}
\end{abstract}


\section{Introduction}

The consideration of large $N$ gauge theories, their
confinement/deconfinement phase-transitions and the possible
relation to the transition from a thermal ensemble to a large
black hole at high-temperatures in the dual string theory is a
topic which is being actively discussed \cite{witten}.  An
important insight \cite{sundborg}--\cite{minwalla} was made by
considering the gauge theories on a compact space, as there is an
additional parameter $R\L$ which can be tuned to weak-coupling
allowing for a perturbation expansion, where $R$ is the size of
the compact space, and $\L$ is the dynamical scale of the gauge
theory.  As the gauge theory is on a compact manifold, the Gauss
law constraint induces interactions between gluons and the matter
multiplets even at zero coupling.

Sundborg \cite{sundborg} and Aharony, {\it et al.}
\cite{aharony,minwalla}, have provided a general framework to
discuss these issues in the context of a unitary matrix model.
This strategy has been developed and extended in a variety of ways
\cite{schnitzer}--\cite{semenoff}. The particular system that
concerns us in this paper is large $N$, $U(N)$ gauge theory with
$N_f$ matter multiplets in the fundamental representation with
$N_f/N$ finite.

We previously have shown \cite{schnitzer} that large $N$,
zero-coupling, $U(N)$ gauge theory with $N_f$ matter fields in the
fundamental representation on a compact spatial manifold $S^{d-1}
\times S^1$, with $(N_f/N)$ finite, undergoes a 3$^{\rm rd}$ order
confinement/deconfinement phase transition at a temperature $T_c$.
The low-temperature phase (phase $A_0$) has free energy of ${\cal
O} (N^2_f)$, interpreted as a gas of color singlet mesons and
glueballs, while the high-temperature phase (phase $A_1$) has a
free energy characterized by $N^2 \: f (N_f/N, \: T)$, interpreted
as that of a gas of gluons and of fundamental and anti-fundamental
matter states.

An intriguing question is the identification of the string dual to
the gauge theory and the associated deconfinement transition.
Witten \cite{witten} and others \cite{sundborg}--\cite{semenoff}
have associated the confinement/deconfinement transition with the
Hawking--Page \cite{buchel} transition of the dual theory. When
$(N_f=0)$ the low-temperature phase has free energy of ${\cal
O}(1)$, while the high temperature phase has free energy of ${\cal
O}(N^2)$, signaling the deconfinement of color singlet glueballs
into gluons.  In the dual string theory this is associated to a
transition from a thermal gas of supergravity modes to a large
black hole \cite{witten}. At zero-coupling of the gauge theory on
$S^{d-1} \times S^1$, this transition is 1$^{\rm st}$ order at the
Hagedorn temperature $T_H$ \cite{sundborg,aharony}.  For pure
Yang--Mills theory on $S^3 \times S^1$, computed to 3-loops in
perturbation this remains a first-order transition
\cite{aharony,minwalla}. At weak-coupling in the gauge theory, on
expects the thermal string/black hole transition to occur for
black holes with size of order of the string scale, while a
comparison with the supergravity approximation requires
strong-coupling in the gauge theory.

When $(N_f/N)$ is finite, at zero-coupling the transition is
3$^{\rm rd}$ order on the gauge side \cite{schnitzer}.  However,
no example is yet known of a string theory with a 3$^{\rm rd}$
order transition to a black hole.  The challenge is to understand
if such a transition exists in the corresponding string dual.  If
the zero-coupling results of a 3$^{\rm rd}$ order transition
survives extrapolation to strong-coupling, then one would have the
(relatively) easier task of finding an appropriate dual in the
supergravity approximation.  This task itself is not without
difficulties.  For example, with $d=4$ one might be tempted to
consider a IIB model with $N \: D3$ branes and $N_f \: D7$ branes,
with $(N_f/N)$ finite in the large $N$ limit.  However, this
simple picture is not possible as the number of allowed $D7$
branes is limited.  One approach to this issue is to consider
$d$=3, with $N \: D2$ branes and $N_f \: D6$ branes at finite
temperature, including back-reaction \cite{cherkis}. Unfortunately
the construction considered in ref.\  \cite{cherkis} only gave the
high-temperature limit of the system, so there was no opportunity
to observe any possible Hawking--Page transition. Another approach
does allow for a back reaction in a $d$=4 string theory with $N_f
\neq 0$ \cite{casero}, but again the putative phase-transition has
not been discussed. Finally, if one has $N_f$ probe $D7$ branes
embedded in $AdS_5 \times S^5$ \cite{mateos,karch}, the
Hawking--Page transition remains first-order even a
strong-coupling but in the probe approximation
\cite{karch,dumitru}. Perhaps this is not surprising as $N_f/N
\rightarrow 0$ in the large $N$ limit in the probe approximation,
so that this should be compatible with $N_f = 0$. Therefore the
possibility of a 3$^{\rm rd}$ order Hawking--Page transition with
$N_f/N$ finite is unresolved.

In this paper we continue the discussion began in ref.
\cite{schnitzer} where $N_f/N$ is finite, but now with
non-vanishing coupling constants.  In Sec. 2 we consider the
planar contribution at large $N$ to $S_{\rm eff}$ (the effective
action) in perturbation theory, as a double expansion in the
't~Hooft coupling $\lambda$ and $N_f/N$.  This is presented in
terms of a matrix model for $U$, given in terms of the zero-mode
of the gauge field

\be
\a = \frac{1}{\o_{d-1}} \int_{S^{d-1}} A_0 \; ,
 \ee

\newpage

 \be
 U = e^{i\b\a} = e^{i\a /T}~~~~~~~
 \ee
 and
 \be
 u_n = \frac{1}{N} \: Tr (U^n)
 \; .
  \ee

We argue in Sec.\ 2 that at low-orders in the double perturbation
theory, there is no solution to the saddle-point equations with
$u_n = 0 \; (n\geq 1)$.  Since the eigenvalue distributions of $U$
can be written as
 \be
 \rho (\th ) = \frac{1}{2\pi} \sum^\infty_{n=-\infty} u_n \:
 e^{in\th}
 \ee
 where $u_n = \bar{u}_n$ and $u_0 =1$, thus there is {\it no}
 phase with $\rho (\th )$ = constant.

In Sec.\ 3 we discuss in detail the phase structure as a function
 of temperature that follows from the $S_{\rm eff}$ of Sec.\ 2 for
 the special case where $(N_f/N) = {\cal O}(\l^2)$.  In this
 example the saddle-point equations require solution of cubic
 equations, and already exhibit many of the main issues of this
 paper.  There is a 3$^{\rm rd}$ order transition from phase $A_0$
to $A_1$ \cite{schnitzer,gross} at a temperature $T_c$ below the
putative Hagedorn temperature, a feature which appears to be a
generic property of  gauge theories with $(N_f/N)$ finite.

In Sec.\ 4 we adapt the strategy of ref. \cite{semenoff} to the
issues of this paper.  In eq'n.\ (4.1) we present the gauge
invariant matrix model consistent with the double perturbation
expansion in $\l$ and $(N_f/N)$   of large $N$ non-Abelian $U(N)$
gauge theory. First we
 recover the zero-coupling results of ref. \cite{schnitzer}.  The
 saddle-point equations which follow from (4.1) are shown in
 general to exclude the phase described by $\rho (\th )$ =
 constant, to any finite order in the double perturbation series
 in $\l$ and $(N_f/N)$.  Thus in this case, the
 low-temperature phase is always in phase $A_0$, {\it i.e.}, that
 described by $\rho (\th ) \neq$ constant, covering $0 \leq \th <
 2\pi$.  We argue, again in perturbation theory, that the
 transition to the single-gapped phase $A_1$ is  3$^{\rm rd}$
 order.

 Sec.\ 5 summarizes our results, and offers plausible conclusions
 regarding the string dual evaluated in the supergravity
 approximation.  It is reasonable to expect that the supergravity
 limit of string theories dual to the class of gauge theories
 considered in this paper exhibit a 3$^{\rm rd}$ order transition
 from a thermal gas of supergravity modes
 to a large black hole.  No such example is  yet known.

 \noindent Issues not discussed in this paper are:\\
 1. ~small black-holes \cite{semenoff,kruczinski},\\
2. ~smoothing out the phase-transitions at finite $N$
 \cite{liu,semenoff}, and\\
3. ~the logical possibility that at a temperature above $T_c$,
 there exist phases $A_n$ such that $\rho (\th )$ is $n$-gapped
 \cite{jurkiewicz}.

  \noindent Since even when $(N_f/N)$ is ${\cal O} (\l^2 )$, {\it c.f.} Sec.\ 3, the
 solution of the saddle-point equations require analysis of a
 cubic, the general analysis of the saddle-point equations can
 involve complicated, model-dependent algebraic equations. The
 principle conclusions of this paper do not depend on such
 considerations.

\renewcommand{\theequation}{2.\arabic{equation}}
\setcounter{equation}{0}

 \section{The Effective Action}

\vspace*{-.1in}

 {\bf A.  Perturbation theory}

 Consider free $U(N)$ gauge theories on $S^{d-1} \times$ time,
 with $N_f$ matter multiplets in the fundamental representation of
 the gauge group with $N_f/N$ finite as $N \rightarrow \infty$.
 The partition function on this compact space (therefore subject
 to Gauss' law constraints) is a matrix model
  \bea
  Z\!\!\!Z (x) & = & \int [dU] \exp \left\{ \sum^\infty_{n=1} \frac{1}{n}
  \; \left[ {\cal Z}_B (x^n) + (-1)^{n+1} {\cal Z}_F (x^n) \right]
 \cdot  \Big[ tr (U^n) + tr (U^{+n})\Big] \right. \\
  & + & \left. \sum^\infty_{n=1} \frac{1}{n} \; \left[ z_B (x^n) +
  (-1)^{n+1} z_F (x^n) \right] tr (U^n) tr (U^{+n}) \right\}
  \nonumber
  \eea
where $z_B (x)$ and $z_F (x)$ are the single-particle partition
functions for the bosons and fermions in the adjoint
representation, while ${\cal Z}_B (x)$ and  ${\cal Z}_F (x)$ are
the single-particle partition functions for the bosons and
fermions in the fundamental representation, where
 \be
 x = e^{-\b} = e^{-1/T}
 \ee
 for $S^{d-1}$ with radius $R$=1.  [$R$ may be restored from
 dimensional considerations as needed.] Recall that
\be
 u_n = \frac{1}{N} \; Tr (U^n)
 \ee

Now consider the same theory with non-vanishing gauge couplings
$g^2$, with $\l = g^2N$ the 't~Hooft coupling.  It is useful, in
this section without a specific theory in mind to keep any scalar
self-couplings and Yukawa couplings at zero. The partition
function can be written as
 \be
 Z\!\!\!Z (x) = \int du_n d\bar{u}_n \: \exp \left[ -
 \tilde{S}_{eff} (u_n)\right] \; .
 \ee
The planar contribution at large $N$ to $S_{eff}$ in perturbation
theory, in a double expansion in $\l$ and $N_f/N$, consistent with
gauge invariance, is
 \bea
 N^{-2} \tilde{S}_{eff} (u_n) & = & S_{eff} (u_n) \nonumber \\
 & = & \Big\{ \sum_n \: G_n (x) |u_n|^2 \nonumber \\
 & + & \l \sum_{m,n} G_{m,n} (x) (u_m u_n u_{-n-m} + c.c.) \nonumber
 \\
& + & \l^2 \sum_{m,n,p} G_{m,n,p} (x) (u_m u_n u_p u_{-n-m-p} +
c.c.) \nonumber
 \\
& + & \ldots \Big\} \nonumber
 \\
 & + & \left( \frac{N_f}{N} \right) \Big\{ \sum_n \: F_n (x) [u_n
 + c.c.]+\lambda\sum_{m,n}F_{m,n}\left[ u_nu_mu_{-n-m}+c.c.\right] \\
 & + & \l^2 \sum_{n,m,p} F_{n,m,p} (x)\left[ u_nu_mu_pu_{-n-m-p}+c.c.\right]  \nonumber
 \\
 & + & \l^3 \sum_{n,m,q,p} F_{n,m,q,p} (x)\left[ u_nu_mu_pu_qu_{-n-m-p-q}+c.c.\right]
  \nonumber
 \\
& + & \ldots \Big\} + {\cal O} \left( \left( \frac{N_f}{N}
\right)^2 \right) \; . \nonumber
 \eea
 Consider the Feynman diagrams of the theory, where each insertion
of a fundamental loop into an adjoint line of a diagram gives a
factor of $\l (N_f/N)$, while a radiative correction by an adjoint
line to a fundamental propagator gives a factor of $\l$.  [Since
we are keeping scalar self-couplings and Yukawa couplings at zero,
these are the only corrections to be considered in this section.]
In (2.5) the leading powers of $\l$ and of $(N_f/N)$ in the
expansion of $\tilde{S}_{eff}$ are exhibited explicitly.    The
various coefficients in (2.5) are a double power series in $\l$
and $(N_f/N)$, beginning at $\l = 0$ and $(N_f/N)=0$.

Near the saddle point of (2.5) one may integrate out the $u_n
(n\geq 2)$, since at the critical point $x_c$ in phase $A_0$,
$|u_n (x_c)|<<|u_1 (x_c)|$ for $n \geq  2$.  Therefore the
effective action for $u_1$ near the saddle-point for $u_1$ is
 \bea
 {N^{-2} \tilde{S}_{eff} (u_1)} & = & \left[ A |u_1 |^2 + B|u_1|^4
 + \ldots \right]  \nonumber
 \\
 & + & \left( \frac{N_f}{N}\right) \Big\{ (u_1 + \bar{u}_1) \left[
 -C + D |u_1|^2 + E |u_1|^4 + \ldots \right]  \nonumber
 \\
& + & \left( u^2_1 + \bar{u}^2_1 \right) \left[ F + G |u_1|^2 + H
|U_1|^4 + \ldots \right]  \nonumber
 \\
& + & \left( u^3_1 + \bar{u}^3_1 \right) \left[ K + L |u_1|^2 + M
|u_1|^4 + \left. \ldots \right] \right\} \nonumber \\
& + & \ldots
 \eea
The leading order contributions to the coefficients in (2.6) are
 \bea
 A & = & {\cal O} (1) \; \; ; \;\; B \; = \; \cO (\l^2 ) \;\; ;
 \nonumber \\
 C & = & {\cal O} (1)  {\rm \; and \; positive \; at \; leading \; order}
 \;\; ; \nonumber \\
 D & = & {\cal O} (\l^2 ) \; \; ; \;\; E \; = \; {\cal O} (\l^4 ) \;\; ;
 \nonumber \\
 F & = & {\cal O} (\l ) \; \; ; \;\; G \; = \; {\cal O} (\l^3 ) \;\; ;
 H \; = \; {\cal O} (\l^5 ) \;\; ; \nonumber \\
 K & = & {\cal O} (\l^2 ) \;\; ; \;  {\rm etc.}
 \eea

 Since each insertion of a fundamental loop in an adjoint line is
 accompanied by a factor of $\l (N_f/N)$, higher-orders in $(N_f/N)$ result in
 higher powers of $\l$ compared to those displayed in (2.6), while
 the ${\cal O} (N_f/N)$ term is that of a single fundamental loop
 with adjoint radiative corrections.  Consider the schematic structure of
 the saddle-point condition which follows from (2.6)
 \bea
 \lefteqn{N^{-2}  \: \frac{\pa \tilde{S}_{eff}}{\pa u_1}} \nonumber \\
 & = & - \left( \frac{N_f}{N} \right) C + u_1 \left\{ 2 B \; \bar{u}^2_1
 + 2  \left( \frac{N_f}{N} \right) F +  \ldots \right\} \\
 & & +  \; \bar{u}_1 \left\{ A + \ldots \right\} + {\cal O} \left( \l
 \left( \frac{N_f}{N} \right)^2 \right) = 0 \nonumber
 \eea
where the curly brackets are a power-series in $u_1$ and
$\bar{u}_1$, with coefficients a series in $\l$.  [We have only
exhibited the leading terms explicitly in the curly brackets.]
Recall that $C$ is a power series in $\l$, with its leading term
${\cal O} (1)$ real and positive, while $A$ has leading-term
${\cal O}(1)$, $F$ of ${\cal O} (\l )$, and $B$ of ${\cal O}
(\l^2)$. Thus, $u_1 = 0$ is {\it not} a solution to (2.8) to the
order we are working. Further, since corrections to the constant
term in (2.8) are a power series in $\l$ and $(\l \: N_f/N)$, the
general structure of the saddle-point equation in perturbation
theory, emphasizing the constant terms, is \pagebreak
 \bea
 \lefteqn{N^{-2}\: \frac{ \pa \tilde{S}_{eff}}{\pa u_1}} \nonumber \\
 & = & - \left( \frac{N_f}{N} \right) C + \sum^\infty_{n=1,\,m=0}
 \l^{n+m}
 \left( \frac{N_f}{N} \right)^{n+1} C_{n,m}  \nonumber\\
 & & +  \; u_1 \left\{ 2B \; \bar{u}^2_1 + \ldots \right\}
 \nonumber\\
& & +  \; \bar{u}_1 \Big\{ A + \ldots \Big\} = 0 \; .
 \eea
The curly brackets represent a perturbative expansion in $u_1$
with the leading term exhibited explicitly, where the $C_n$ are
independent of $u_1$ and $\bar{u}_1$, but a power-series in $\l$.
Since $C = {\cal O} (1)$, and positive at leading order, {\it cf.}
(2.7), $u_1 = 0$ is {\it not} a solution to (2.9) in any finite
order of the double perturbation theory, as the powers of $\l$ and
$(N_f/N)$ do not match. A solution $u_1=0$ to (2.9) would require
 \be
 \sum^\infty_{n=1,\,m=0} \l^{n+m} \left( \frac{N_f}{N} \right)^n
 C_{n,m}
 \stackrel{?}{=} C
 \ee
which is not possible in any finite order of the double expansion,
since $C$ at leading order is independent of $\l$ and $(N_f/N)$.
If (2.10) has a solution, it must be a non-perturbative or
all-orders result. As a consequence, $u_1=0$ does {\it not} solve
the saddle-point equations in perturbation theory if $(N_f/N)$
finite is considered as an expansion parameter of the theory.

\noindent{\bf B. Phases}

The eigenvalue distribution of the matrix $U$ can be written as
 \be
\rho (\th ) = \frac{1}{2\pi} \sum^\infty_{n=-\infty} u_n \:
e^{in\th}
 \ee
where $u_{-n} = \bar{u}_n$ and $u_0 = 1$.  The $u_n$, with $n \geq
2$ can be expressed in terms of $u_1$ near the saddle-point by
means of the saddle-point equations that follow from (2.5), which
is how (2.6) is obtained.  Schematically, (not being explicit
about coefficients and not distinguishing between $u_n$ and
$u_{-n}$), one has
  \bea
  u_2 & \sim & \left(\frac{N_f}{N} \right) + \l \left(\frac{N_f}{N} \right)
  u_1 + \l \: u^2_1 + \ldots \nonumber
  \\
 u_3 & \sim & \left(\frac{N_f}{N} \right) + \l
  u_1u_2+\lambda\left({N_f\over N}\right)u_1  + \ldots\nonumber
  \\
& \sim & \left(\frac{N_f}{N} \right) + \l \left(\frac{N_f}{N}
\right)
  u_1 + \l^2  \left(\frac{N_f}{N} \right)  \: u^2_1 \nonumber
  \\
 &&  + \; \l^2 u^3_1 + \ldots \; ,
 \eea
which displays the leading orders in $\l$ and $(N_f/N)$ in (2.12),
and similarly with its generalization to $u_n$. Eqn.\ (2.12)
reiterates that $u_n \neq 0$ are the only solutions to the
saddle-point equations within the context of perturbation theory.

Thus, for $(N_f/N)$ finite in the large $N$ limit, (2.11)--(2.12)
and the saddle-point equations require, in the double perturbation
expansion that
 \be
 \rho (\th ) \neq {\rm \; constant \; .}
 \ee
 Therefore, the allowed phases (in the notation of ref. \cite{schnitzer})
 are
 \be
 (i) ~~~ (A_0): \;\; \rho (\th ) \;\;\; {\rm completely \; covers
 \; the \; circle, \; but \; is \; not \; constant.}
 \ee
$$
(ii) ~~~ (A_1): \;\; \rho (\th ) =0 \hspace*{3.15in}
$$
for a single finite gap in the distribution.  The phase $(A_0)$ is
that described by the saddle-point, while $(A_1)$ can be described
by the methods of refs.\
\cite{sundborg,aharony,schnitzer,semenoff,gross}. The criterion
that $\rho (\th ) \geq 0$ implies from (2.12)--(2.13) that
 \be
 |u_1| \leq \frac{1}{2} + {\cal O}\left(\frac{N_f}{N}\right) + {\cal
 O}(\l ) + {\cal O} \left( \l \: \frac{N_f}{N}\right) + \ldots
 \ee
 in the double perturbation theory.  The $(A_0,A_1)$ phase
 boundary occurs when $\rho (\th ) = 0$ at a single point, which
 then implies that (2.15) is an equality.  The transition to phase
 $A_1$ is third-order, as can be deduced from an analysis
 analogous to that of ref. \cite{schnitzer}.  [It is a logical
 possibility \cite{jurkiewicz} that there could be additional phases $A_n$, with $n$
 finite gaps $(n \geq 2)$, whose phase boundary occurs at
 temperatures above that of the $(A_0,A_1)$ transition, but this
 is outside the scope of our discussion.  Further, this does not
 occur in theories with $\l = 0$, $(N_f/N) \neq 0$ \cite{schnitzer}.]

 \section{Phase Structure of an Example}

\renewcommand{\theequation}{3.\arabic{equation}}
\setcounter{equation}{0}

In this section we discuss in detail the phase structure for the
case when $(N_f/N) = \cO (\l^2)$.  The purpose of the example is
to illustrate a detailed discussion of the saddle-point equations,
and evolution of the phase-structure with temperature.  Although
there are only two phases in this case, the details of the
evolution with temperature depend on the relative magnitudes of
parameters.  When $(N_f/N) = \cO (\l^2)$, insertions of
fundamental loops into the diagrams of pure Yang--Mills theory can
be neglected to the order in $\l^2$ we are considering.  Further,
this example is interesting, as one can additively combine the
contribution of fundamental matter to $\tilde{S}_{eff}$ with the
results of ref. \cite{minwalla}, where pure Yang--Mills theory in
$d$=4 was analyzed to three-loop order.

\noindent{\bf A. ~General considerations}

Consider (2.6)--(2.7) correct to $\cO (\l^2)$, with  $(N_f/N) =
\cO (\l^2)$.  Only the first three terms of (2.6) have
coefficients of that order so that to leading order
 \bea
 N^{-2} \tilde{S}_{eff} (u_1) & = & m^2_1 |u_1|^2 + b|u_1|^4 \nonumber \\
 && - d(u_1 + \bar{u}_1)
 \eea
 where
 $$
 m^2_1 = K(T_H - T) \; ,
 $$
 with
 \be
  K = \cO (1)\; ,
  \ee
  $b =\cO (\l^2)$, $d > 0$ and $\cO (N_f/N)$, $(N_f/N) = \cO (\l^2)$, and where
 $T_H$ is defined as the temperature where $m^2_1 = 0$.  Separate
 $u_1 = \tilde{u}_1 + i\tilde{u}_1$ into real and imaginary parts,
 so that the saddle-point equations that follow from (3.1) are
 \bea
 m^2_1 \tilde{u}_1 + 2b \: \tilde{u}_1 \: | u_1 |^2 - d  & = & 0 \nonumber
 \\
m^2_1 \hat{u}_1 + 2b \: \hat{u}_1 \: | u_1 |^2 & = & 0
 \eea
which requires $\hat{u}_1 = 0$; consistent with $d =(N_f/N) C$
real.  To the order we are considering, $C$ is computed at $\l =
0$, so that it has the same value as in ref. \cite{schnitzer}. The
restriction (2.15) now becomes $|u_1| \leq 1/2 + \cO(\l)$ in phase
$A_0$.

In order to study phase $A_0$, we need the saddle-point solution
to
 \be
 N^{-2} \tilde{S}_{eff} (u_1) = m^2_1 \; | u_1|^2 + b\; |u_1|^4 - 2 \; d \; u_1
 \ee
 which requires
 \be
 2b \; u^3_1 + m^2_1 \; u_1 - d = 0 \; .
 \ee
 Solutions to (3.5) are characterized by the discriminant
 \be
 D = -\frac{1}{2} \left( \frac{m^2_1}{b} \right)^3 - 27
 \left(\frac{d}{2b}\right)^2 \; .
 \ee
 When $D > 0$ there are 3 real roots from (3.5), while if $D<0$
 there is 1 real root, and 2 complex roots which are not relevant,
 as $u_1$ is real.  If
 \be
 \left| \: \frac{d}{2b} \: \right|^{2/3} > >
 \left| \: \frac{m^2_1}{3b} \: \right|
\ee then $D < 0$.

\noindent If
 \be
 \left| \: \frac{d}{2b} \: \right|^{2/3} < <
 \left| \: \frac{m^2_1}{3b} \: \right| \; ,
\ee the sign of $D$ depends on the sign of $(m^2_1 / b)$, in which
case the solution for $u_1$ is approximately that of ref.
\cite{aharony,minwalla} {\it except} instead of $u_1 = 0$, now
$$
u_1 \simeq (d/m^2_1 ) > 0 \;\;\;\;\; {\rm for} \; m^2_1 > 0 \; .
$$
Since $(d/2b) \sim \cO (1)$ in this section, (3.8) always occurs
for $T$ sufficiently below $T_H$.  For example, $m^2_1 \sim \cO
(\l )$ is sufficient for this to occur.

In the next subsection we concentrate on a toy model, with $b <
0$, recalling that $b < 0$ for pure Yang--Mills theory in $d$=4
\cite{aharony}.  The analysis for $b > 0$ is qualitatively
similar,  which we do not present here.

\noindent{\bf B. ~Toy model}

In the general case, $u_n \neq 0$ for $n\geq 1$, which leads to
the constraint (2.15).  In this section $(N_f/N) \sim \cO (\l^2)$,
so that
$$
|u_1| \leq 1/2 + \cO (\l )
$$
in the saddle-point phase $A_0$.  One can simplify the analysis
without sacrificing any of the qualitative features of the
physics, so that for the remainder of this section we consider
(3.4), but with $u_n = 0$ for $n \geq 2$.  Then $\rho (\th ) =
1/2\pi \:(1 + 2u_1 \cos \th )$, which implies that
 \be
 | u_1| \leq 1/2
 \ee
 with the $(A_0 , \; A_1)$ phase boundary occurring when $|u_1|$
 increases to $|u_1| = 1/2$ in the $A_0$ phase, with the
 transition to $A_1$ of 3$^{\rm rd}$ order.

\noindent{\bf C. ~{\bf\it b} $<$ 0}

We track the evolution of the system for increasing $T$, beginning
with $T<T_H$.

\noindent (i) ~$m^2_1 > 0 \; ; \; D > 0$

If $D >0$, this requires
\be
|b| < \frac{2}{27} \; \frac{(m^2_1)^3}{d^2}
 \ee
 and an approximate solution to (3.5) is
 \bea
 u_1 & \simeq & d/m^2_1 \nonumber \\
&& < \left[ \frac{2}{27} \; \frac{d}{|b|} \right]^{1/3} \; .
 \eea
[The two other real solutions to (3.5) are that of maxima of
$\tilde{S}_{eff} (u_1)$, and thus not relevant.]  For sufficiently
large $m^2_1 \; [m^2_1 \sim \cO (\l )$ is sufficient], $|u_1 | <
1/2$ so that the system is in phase $A_0$, and
 \be
 N^{-2} \tilde{S}_{eff} (u_1) \simeq -d^2/m^2_1  \; .
 \ee
 The free-energy is
 \be
F/T =\tilde{S}_{eff} (u_1)
 \ee
so that
 \bea
 F/T & = & - N^2 (d^2/m^2_1 ) \nonumber \\
 & \simeq & \cO (N^2_f )/(T_H-T) \; ,
 \eea
 since $d \sim \cO (N_f/N)$.  Thus (3.14) has the same form as
 (1.1) of ref. \cite{schnitzer}, and the interpretation of phase $A_0$ as
 that of a gas of gluons and color singlet mesons.  Therefore for
 sufficiently low temperatures one is in phase $A_0$.

The saddle-point value of $u_1$ increases as $T$ increases.
Consider

\noindent (ii) ~$D > 0$, near $D=0$; $m^2_1 > 0$

Using (3.6) we have
 \be
 m^2_1 = \left[ \frac{27}{2} \: d^2 \; |b| \right]^{1/3} + \d \: m^2
 \; ,
 \ee
 and
 \bea
 u_1 & = & \left[ \frac{d}{4|b|} \right]^{1/3} \left\{ 1 - \cO
 \left(\frac{\d m^2}{m^2_1} \right)\right\} \nonumber \\
 & = & \frac{d}{2m^2_1} \left\{1 -  \cO
 \left(\frac{\d m^2}{m^2_1} \right)\right\} \; .
 \eea
We verify that $\frac{\pa^2 \tilde{S}_{eff}}{\pa^2u_1} \: (u_1)
> 0$, so that this minimum is stable.

If $u_1 < 1/2$, the system is still in phase $A_0$ at this
temperature, while if $u_1 > 1/2$ the system has already made a
transition to phase $A_1$ at a lower-temperature.

\noindent (iii) ~$D = 0$; $m^2_1 > 0$
\bea
u_1 & = & \left[ \frac{d}{4|b|} \right]^{1/3} \nonumber \\[.1in]
& = & \frac{d}{2(m^2_1)} \; .
 \eea
However,
\be
\frac{\pa^2 \tilde{S}_{eff}}{\pa u^2_1} \: (u_1) = 0
 \ee
and
\be
\frac{\pa^3 \tilde{S}_{eff}}{\pa u^3_1} \: (u_1) = - |b| \;u_1 < 0
 \ee
so that (3.17) is not a stable minimum.  The system evolves to
$|u_1| = 1/2$ at which value it makes a transition to phase $A_1$,
{\it if} it has not already made this transition at a lower
temperature.  [See the discussion for (ii) above.]

\noindent (iv)  ~$m^2_1 \leq 0$  ; $D < 0$

The solution to (3.5) is a single maximum at $u_1 < 0$, and not a
minimum.  There is no stable saddle-point for $|u_1| \leq 1/2$, so
that at $T = T_H$ ({\it i.e.}, $m^2_1 = 0$), the system is already
in phase $A_1$.

In summary, for $(N_f/N) \sim \cO (\l^2)$ and $T$ sufficiently
small, the system is in phase $A_0$.   At some $T_c < T_H$ there
is a third-order phase transition to the high temperature phase
$A_1$.  The free energy in phase $A_0$ behaves as
 \be
 F/T \sim N_f^2 f_1 (T) \hspace{.75in} 0 \leq T \leq T_c
 \ee
and in phase $A_1$
 \be
 F/T \sim N^2 f_2 \left( \frac{N_f}{N} \; , \; T \right) \hspace{.75in}  T \geq
 T_c\; .
 \ee
The low-temperature phase is interpreted as a gas of (color
singlet) mesons and glueballs, which makes a third-order
confinement-deconfinement transition to a gas of gluons and
fundamental and anti-fundamental matter states at high
temperatures.

 \section{Generalization}

\renewcommand{\theequation}{4.\arabic{equation}}
\setcounter{equation}{0}

We use the strategy of ref.\ \cite{semenoff} to discuss the issues
of this paper suitable for all orders of perturbation theory.  In
doing so, we modify their set-up to suit our problem.  Recall that
$u_n =1/N \: Tr \: U^n$, so that (2.5) is a special case of
\pagebreak
 \bea
 S(U,U^+) & = & \sum^\infty_{n=1} \frac{a_n (x)}{n} \; (Tr \: U^n)
 (Tr \: U^{+n}) \nonumber \\[.1in]
 & + & N_f  \sum^\infty_{n=1} \left[ \frac{b_n (x)}{n} \; (Tr \: U^n)
 + c.c.\right] \nonumber \\[.1in]
& + & N^2 \: \sum_{k,k^\prime} \a_{\bar{k},\bar{k}^\prime} (x) \;
Y_{\bar{k}} (U) Y_{\bar{k}^\prime} (U^+) \nonumber\\[.1in]
& + & N \: \left(\frac{N_f}{N} \right) \left[
\sum_{\bar{k},\bar{k}^\prime} \b_{\bar{k},\bar{k}^\prime}(x)\;
Y_{\bar{k}}(U)Y_{\bar{k}^\prime}(U^+) \right]
\sum^\infty_{n=1}\Big[ (Tr \: U^n) + (Tr \: U^{+n})\Big]
 \eea
where
 \be
 Y_{\bar k} (U) = \prod^\infty_{j=0} (Tr (U/N)^j)^{k{_j}}
 \ee
 and where $\bar{k}$ and $\bar{k}^\prime$ are arbitrary vectors of
 non-negative entries.  Reality of the action and perturbation
 theory require \cite{semenoff}
\be
\a_{\bar{k},\bar{k}^\prime} = \a^*_{\bar{k}^\prime , \bar{k}} =
\a_{\bar{k}^\prime ,\bar{k}}
 \ee
and similarly for $\b_{\bar{k},\bar{k}^\prime}$.  It is useful to
define
 \be
 |\bar{k}| = \sum_j \: k_j
 \ee

There is considerable redundancy in (4.1) which we remove in part by
setting
 \be
 \begin{array}{rcl}
 \a_{\bar{k},\bar{k}} = 0 & {\rm for} & |\bar{k} | = 1 \nonumber
 \\
\a_{\bar{k},0} = 0 & {\rm for} & |\bar{k} | = 1 \nonumber
 \\
\b_{0,0} = 0  & &
\end{array}
\ee

Contact can be made with the formulation of (2.5) by noting that
the leading behavior in perturbation theory of the coefficients in
(3.1) are
 \bea
 a_n (x) & = & \cO (1) \nonumber \\
 b_n (x) & = & \cO (1) \nonumber \\
\a_{\bar{k},\bar{k}^\prime} & \sim & ( \l )^{|\bar{k} |
 + |\bar{k}^\prime |-2} \nonumber \\
\b_{\bar{k},\bar{k}^\prime} & \sim & (\l )^{|\bar{k} |
 + |\bar{k}^\prime |} \; .
 \eea
Beyond the leading order of (4.6), the coefficients are a power
series in $\l$ and $(N_f/N)$, as well as any other coupling
constants present in the problem.

We take advantage of the Gaussian trick extensively exploited in
ref. \cite{semenoff} to write \pagebreak
 \bea
 \lefteqn{\exp \left\{ \sum^p_{n=1} \frac{a_n}{n} \; tr (U^n) tr
 (U^{+n}) \right. }
 \nonumber \\
& &  + \;  \left. N_f \sum^p_{n=1} \frac{b_n}{n}\; \Big[ tr (U^n)
+ tr
(U^{+n}) \Big] \right\} \nonumber \\
& = & \left(\frac{N^2}{2\pi}\right)^p \int \prod^p_{n=1}
\left(\frac{a_n}{n}\right) dg_n d\bar{g}_n \: \exp -N^2
\sum^p_{n=1}\left(\frac{a_n}{n}\right) g_n \bar{g}_n
\nonumber \\
&& \cdot \; \exp N \sum^p_{n=1} \frac{1}{2n} \; \Big[c_n tr (U^n)
+ \bar{c}_n tr (U^{+n})\Big]
 \eea
where
 \be
 c_n = 2 \left[ a_n \; g_n + \frac{N_f}{N} \: b_n \right] \; .
 \ee
Eq'ns. (4.7)--(4.8) are closely related to (3.1) and (3.5) of
ref.\ \cite{schnitzer}, except that now we allow $a_n$ and $b_n$
to be power series in $(N_f/N)$ and $\l$, with the leading order
of $a_n$ and $b_n$ of $\cO (1)$, given by the zero-coupling
expressions of (2.1).

Define the free-energy $F(g_n, \bar{g}_n)$ by
 \bea
\lefteqn{\exp \Big[ N^2 F(g_n, \bar{g}_n) \Big]} \nonumber \\
& = & \int [dU] \exp \left\{ N \sum^\infty_{n=1} \frac{c_n}{2n} \;
\Big[ tr(U^n ) + tr (U^{+n}) \Big] \right\}
 \eea
 which implies that
 \be
 F(g_n,\bar{g}_n) = \frac{1}{4} \sum^\infty_{n=1} \frac{1}{n} \;
 c_n \bar{c}_n
 \ee
in phase $A_0$.

\noindent A.~  The Free-theory Revisited

It is instructive to consider further (4.1), but first with
$\a_{\bar{k},\bar{k}^\prime} = \b_{\bar{k},\bar{k}^\prime} = 0$.
Using the Gaussian identity \cite{semenoff}
 \bea
 \lefteqn{ \exp \left[ -N^2 \sum^p_{n=1} \left(\frac{a_n}{n} \right)
 g_n\bar{g}_n \right]}   \nonumber \\
& = & \left( \frac{N^2}{\pi} \right)^p \int \prod^p_{n=1}
\left(\frac{a_n}{n} \right) d\m_n d\bar{\m}_n \nonumber \\
&& \cdot \; \exp \left\{ -N^2 \sum^p_{n=1} \left(\frac{a_n}{n}
\right)
  \m_n \bar{\m}_n + i N^2\left(\frac{a_n}{n} \right)  [\m_n \bar{g}_n
  + \bar{\m}_n g_n] \right\}  \; .
  \eea
The partition function is
 \be
 Z\!\!\!Z = \left( \frac{N^4}{2\pi^2}\right)^p \int \prod^p_{n=1}
 dg_nd\bar{g}_n d\m_nd\bar{\m}_n \exp (N^2 S_{eff})
 \ee
 where here
 \bea
 S_{eff} & = & -\sum^p_{n=1} \left( \frac{a_n}{n} \right)
 \m_n\bar{\m}_n + i \sum^p_{n=1}\left( \frac{a_n}{n} \right)
 [\m_n \; \bar{g}_n + \bar{\m}_n \; g_n] \nonumber \\
 && + \; F(g_k,\bar{g}_k) \; .
 \eea
The saddle-point equations are
 \bea
\frac{\pa S_{eff}}{\pa g_k} & = & i\bar{\m}_k + \left[ a_k \;
\bar{g}_k
+ \left(\frac{N_f}{N} \right) b_k \right] = 0 \; ; \\
[.2in] \frac{\pa S_{eff}}{\pa \m_k} & = & - \;
\left(\frac{a_k}{k}\right) \left[ \bar{\m}_k + i\bar{g}_k \right]
= 0
 \eea
and two analogous equations for variations with respect to
$\bar{g}_k$ and $\bar{\m}_k$.  Since $a_k (x) \neq 0$, one has the
solution
 \be
 g_k = \bar{g}_k = \frac{(N_f/N)b_k}{(1-a_k)}
 \ee
 since $a_k(x)$ and $b_k(x)$ are real.  The density of eigenvalues
 in the phase $A_0$ is
  \be
  \rho (\th ) = \frac{1}{2\pi} \left\{ 1 + \sum_{k=1} \Big[ g_k \exp
  (ik\th ) + \bar{g}_k \exp (-i \: k\th ) \Big]\right\} \; ,
  \ee
so that from (4.8) and (4.16), one has
 \be
 \rho_k = g_k = c_k/2 \; .
 \ee
Thus, (4.16) is identical to eq'ns.\ (3.9) of ref.
\cite{schnitzer}, where there is a 3$^{\rm rd}$ order GWW
phase-transition \cite{gross} to phase $A_1$, interpreted as a
confinement-deconfinement transition. Note that since $(N_f/N)
\neq 0$, $g_k = 0$ is {\it not} a possible solution, so that the
phase $\rho (\th )$ = constant does not occur.  Further discussion
of the consequences of (4.9) and computation of $\rho (\th )$ for
phase $A_1$ may be found in ref.\ \cite{schnitzer}.

Thus, we have verified that the above formalism, with
$\a_{\bar{k},\bar k^\prime} = \b_{\bar k, \bar k^\prime} = 0$,
reproduces ref.\ \cite{schnitzer}.  We now return to the general
case described by eq'ns.\ (4.1)--(4.6).

\noindent A.~  The General Case

Equation (4.13) generalizes to
 \bea
 S_{eff} & = & - \sum_{n=1} \left( \frac{a_n}{n} \right) \m_n \bar
 \m_n \nonumber \\
 & + & i \sum_{n=1} \left( \frac{a_n}{n} \right) [\m_n \bar g_n +
 \bar \m_n g_n ] \;\; - F (g_k, \bar g_k ) \nonumber \\
 & + & \sum_{\bar k,\bar k^\prime} \a_{\bar k,\bar k^\prime} (-i )
^{|k|+|k^\prime |} Y_{\bar k} (\bar \m) Y_{\bar k^\prime} (\m)
\nonumber \\
& + & \left(\frac{N_f}{N}\right) \sum_{\bar k,\bar k^\prime}
\b_{\bar k,\bar k^\prime} (-i ) ^{|\bar k|+|\bar k^\prime |} \;
Y_{\bar k} (\bar \m) Y_{\bar k^\prime} (\m) \sum_{n=1} (-i)^n
[\m_n + \bar \m_n ]
 \eea
where
 $$
 Y_{\bar k} (\bar \m)  =  \sum_{j=1} (\bar \m_j )^{k_j}
 $$
 and
  \be
  Y_{\bar k{^\prime}} (\m )  =  \sum_{j=1} (\m_j )^{k_j^\prime}
  \ee

The saddle-point equations at large $N$ that follow from (4.19)
are
 \be
 \frac{\pa S_{eff}}{\pa g_k} = \left( \frac{a_k}{k} \right) \left\{
 i \: \bar \m_k + \left[ a_k \; \bar g_k + \left( \frac{N_f}{N}
 \right) b_k \right] \right\} = 0
 \ee
\bea
 \frac{\pa S_{eff}}{\pa \m_j} & = & - \left( \frac{a_j}{j} \right)
 \Big[ \bar \m_j + i \: \bar g_j \Big] \nonumber \\
 & + & \sum_{\bar k , \bar  k^\prime}\a_{k,k^\prime}(-i)^{|k|+|k^\prime|} \:
 \frac{(k_j^\prime )}{\m_j} \: Y_{\bar k} (\bar \m) Y_{\bar k^\prime}
 (\m ) \nonumber \\
 & + & \left( \frac{N_f}{N} \right)
 \sum_{\bar k , \bar  k^\prime}\b_{\bar k,\bar k^\prime}(-i)^{|\bar k|+|\bar k^\prime|} \:
 \frac{(k_j^\prime )}{\m_j} \: Y_{\bar k} (\bar \m) Y_{\bar k^\prime} (\m )
 \sum_{n=1} (-i)^n (\m_n +\bar \m_n)
 \nonumber \\
& + & \left( \frac{N_f}{N} \right)
 \sum_{\bar k , \bar  k^\prime}\b_{k, k^\prime}(-i)^{|\bar k|+|\bar k^\prime|} \:
Y_{\bar k} (\bar \m) Y_{\bar k^\prime} (\m)
 (-i)^j =0
 \eea
 and two analogous equations for $\pa S_{eff}/\pa\bar g_j$ and
 $\pa S_{eff}/\pa\bar \m_j$.  Since $a_j \neq 0$, (4.21) implies
 that
 \be
 \bar \m_j = i \left[ a_j \: \bar g_j + \left( \frac{N_f}{N}
 \right) b_j \right] \; ,
 \ee
 and similarly $\m_j = i [a_j \; g_j + (N_f/N) b_j]$.  One then
 substitutes (4.23) into (4.22) to obtain an equation for $\bar
 g_j$  in terms of $a_j, \; b_j, \; \a_{\bar k,\bar k^\prime}$ and
$\b_{\bar k,\bar k^\prime}$, and analogously for $g_j$.  The
solution to (4.22)--(4.23) for $\bar g_j$ and its analogue for
$g_j$ determines the distribution $\rho (\th )$ in the ungapped
phase, via (4.17) which is still valid.

We now ask whether $\rho (\th )$ = constant is possible when
$(N_f/N) \neq 0$.  If so, this would require that $\bar g_j = 0$
for all $j$ in (4.22)--(4.23), and analogously for $g_j$.  If
$k^\prime_j \neq 0$, $g_j = \bar g_j = 0$ would require
 \bea
 \left( \frac{a_j}{j} \right) b_j & \stackrel{?}{=}& \left\{
 \sum_{|\bar k |,|\bar k^\prime | } \a_{\bar k, \bar
 k^\prime} \left( \frac{k^\prime_j}{b_j} \right) \left(
 \frac{N_f}{N} \right)^{|\bar k |+|\bar k^\prime | - 2} \right. \nonumber
 \\
 & - & 2 \sum_{|\bar k |,|\bar k^\prime | } \b_{\bar k,
 \bar k^\prime}\left( \frac{k^\prime_j}{b_j} \right) \left(
 \frac{N_f}{N} \right)^{|\bar k |+|\bar k^\prime | } \left[ \sum_n
 (-1)^{n+1} b_n \right] \nonumber
 \\
 & - & \left. i \sum_{|\bar k |,|\bar k^\prime | } \b_{\bar k,
 \bar k^\prime} \left( \frac{N_f}{N} \right)^{|\bar k |+|\bar k^\prime |}
  (-i)^{j+1} \right\} Y_{\bar k} (b) Y_{\bar k^\prime} (b) \; .
  \eea
Note that $a_j$ and $b_j$ are both non-zero and of $\cO (1)$ in
$(N_f/N)$, while the right-hand side of (4.24) is a power series
in $(N_f/N)$, beginning with leading terms of $\cO (N_f/N)$ [see
(4.5)].  Thus to any finite order in $(N_f/N)$ (4.24) can not be
satisfied by matching powers of $(N_f/N)$. One concludes that
\linebreak $\rho (\th )$ = constant is excluded in any finite
order in the double perturbation theory. At low temperatures the
system is in phase $A_0$, and is {\it not} described by $\rho (\th
)$ = constant at any temperature.

The transition from phase $A_0$ to $A_1$ is driven by the
third-order phase transition in $F (g_k, \bar g_k)$, {\it c.f.}
(4.10) and (4.13), which occurs when $\rho (\pi ) = 0$.  This
occurs when
 \be
 g_1 = \bar g_1 \simeq 1/2 \; .
 \ee
[Equality occurs if one may neglect $g_k$ for $k \geq 2$.  See
Sec.\ 3 for example.]  Note that $a_j (x)$ and  $b_j (x)$ are
continuous functions of $x$ as one passes through the $(A_0,A_1)$
phase transition.  From (4.23) this implies that $\bar \m_j$ and
$\bar g_j (x)$ [and similarly $\m_j (x)$ and $g_j (x)$] are
smoothly related to each other at the ($A_0$, $A_1$) boundary.
Equation (4.19) can be written as
 \be
 S_{eff} = - F (g_k, \bar g_k) + f(g_k, \bar g_k)
 \ee
where $F(g_k, \bar g_k)$ has a third-order transition at the
$(A_0, A_1)$ boundary.  Therefore it is plausible to assume that
$f (g_k, \bar g_k )$ is smooth, with continuous derivatives at the
phase-boundary. In the absence of an ``accidental" third-order
transition in $f(g_k, \bar g_k)$ which cancels that of $F(g_k,
\bar g_k)$, or a lower-order discontinuity which seems unlikely
given the smooth structure of $f(g_k, \bar g_k)$, $S_{eff}$ has a
third-order phase transition at the $(A_0,A_1)$ boundary.

\section{Conclusions}

The phases of large $N$ gauge theory at finite temperature on
$S^{d-1} \times$ time are described by $\rho (\th )$, the density
of the eigenvalues of $U$ in the unitary matrix model.  When
$(N_f/N)$ is finite in the large $N$ limit, $\rho (\th )$ =
constant is {\it not} a solution of the saddle-point equations at
any finite order in the double perturbation theory.  The
saddle-point equations describe a phase $A_0$, where $\rho (\th )
\neq $ constant, but covers the interval $0 \leq \th < 2\pi$.
There is a transition to a second phase $A_1$, where $\rho (\th )$
is single gapped; the transition taking place at a critical
temperature $T_c$, which is lower than a putative Hagedorn
temperature.  The above conclusions are correct to any finite
order in a double perturbation series in the 't~Hooft coupling $\l
$ and in $(N_f/N)$.  In order for the low-temperature phase {\it
not} to be $A_0$, one presumably would require a phase-transition
as one increased $(N_f/N)$ at fixed $\l$, or increased $\l$ at
fixed $(N_f/N)$, keeping the temperature fixed.  We speculate that
this is not the case.

It is a reasonable assumption that generically $f(g_k, \bar g_k)$,
{\it c.f.} (4.26), is sufficiently smooth that the $(A_0, A_1)$
phase transition is controlled by $F(g_k, \bar g_k)$, {\it c.f.}
(4.10) and (4.26), and is therefore 3$^{\rm rd}$ order.  It is
plausible that these results, valid in any finite order of the
double perturbation theory, should also hold at a strong coupling
$\lambda$ satisfying $N_f/N<1/\lambda$ as well. Gravitational
approximations are valid at large $g_sN$, so that the backreaction
on D branes is small. To keep open string perturbation theory in
control, one also needs $g_sN_f\ll 1$ which means
$N_f/N\ll1/\lambda$ in the large $N$ limit. We conjecture that the
third order phase transition is valid if these conditions are
meet.\footnote{We thank the referee for this argument.} This would
then imply that the string theory dual to the gauge theory would
exhibit a 3$^{\rm rd}$ order Hawking--Page transition (dual to the
confinement/deconfinement transition) in the supergravity limit of
the string theory.  No such transition has yet been found when $(N_f
/N)$ = finite in the large $N$ limit, as one must go beyond the
probe approximation, and consider the back-reaction of the flavor
branes.  There have been preliminary studies involving the
back-reaction of flavor branes \cite{cherkis,casero}, but are not
able to discuss the Hawking--Page phase transition from a thermal
ensemble to a large black hole. Discovery of such a 3$^{\rm rd}$
order transition in the string dual to a large $N$ gauge theory with
$(N_f /N) \neq 0$, would be an exciting confirmation of duality in
the context of the ideas of this paper.

\noindent{\bf Acknowledgement}

We thank Albion Lawrence for his comments and for reading the
manuscript, and the referee for important suggestions for improving
the paper.

 \end{document}